\documentclass{new_tlp}
\usepackage{mathptmx}
\usepackage{color}

\newtheorem{example}{Example}

\newcommand{\tuple}[1]{\ensuremath{\langle{#1}\rangle}\xspace}

\usepackage{amsmath}
\usepackage{listings}
\usepackage{todonotes}
\usepackage{xspace}
\usepackage{float}
\usepackage{amssymb}

\makeatletter
\lst@Key{countblanklines}{true}[t]{\lstKV@SetIf{#1}\lst@ifcountblanklines}
\lst@AddToHook{OnEmptyLine}{%
    \lst@ifnumberblanklines\else%
    \lst@ifcountblanklines\else%
    \advance\c@lstnumber-\@ne\relax%
    \fi%
    \fi}
\makeatother
\lstdefinelanguage{asp}{
    breakatwhitespace=true,
    captionpos=b,
    numbers=left,
    numbersep=5pt,
    numberblanklines=false,
    countblanklines=false,
    commentstyle=\colour{gray},
    frame=bt, framexbottommargin=5pt, framextopmargin=5pt,
    aboveskip=5pt, belowskip=5pt,
    abovecaptionskip=10pt,
    basicstyle=\small\ttfamily
}
\lstnewenvironment{asp}{
    \lstset{
        language=asp,
        showstringspaces=false,
        keepspaces=true,
        formfeed=\newpage,
        tabsize=4,
        numbers=none,
        breaklines=true,
        literate={~} {$\sim$}{1},
        frame=none,
    }
}{}

\usepackage{environ}
\usepackage{tikz}
\usepackage{url}
  \title[ASP Chef grows Mustache to look better]
        {ASP Chef grows Mustache to look better}

  \author[M. Alviano, W. Faber, L.A. Rodriguez~Reiners]
         {MARIO ALVIANO\\
         University of Calabria, Italy\\
         \email{mario.alviano@unical.it}
         \and 
         WOLFGANG FABER\\
         University of Klagenfurt, Austria\\
         \email{Wolfgang.Faber@aau.at}
         \and 
         LUIS~ANGEL RODRIGUEZ~REINERS\\
         University of Calabria, Italy\\
         \email{luis.reiners@unical.it}
         }

\jdate{March 2003}
\pubyear{2003}
\pagerange{\pageref{firstpage}--\pageref{lastpage}}
\doi{XXX}

\begin{document}

\label{firstpage}

\maketitle

  \begin{abstract}
We present ASP Chef Mustache, an extension of ASP Chef that enhances template-based rendering of ASP solutions using a logic-less templating system inspired by Mustache.
Our approach integrates data visualization frameworks such as Tabulator, Chart.js, and vis.js, enabling interactive representations of ASP interpretations as tables, charts, and graphs. 
Mustache queries in templates support advanced constructs for formatting, sorting, and multi-stage expansion, facilitating the generation of rich, structured outputs. 
We demonstrate the power of this framework through a series of use cases, including data analysis for the Italian VQR, visualization of blocking sets in graphs, and scheduling problems. 
The result is a versatile tool for bridging declarative problem solving and modern web-based visual analytics.
%
  \end{abstract}



\section{Introduction}\label{sec:intro}

Answer Set Programming (ASP) is a well-established declarative paradigm for solving complex combinatorial problems, offering a clean separation between problem modeling and computation \cite{DBLP:journals/cacm/BrewkaET11,DBLP:journals/aim/ErdemGL16,DBLP:books/sp/Lifschitz19,DBLP:journals/tplp/KaminskiRSW23,DBLP:journals/ai/AlvianoDFPR23}. 
While modern ASP solvers can efficiently compute answer sets, interpreting and presenting these solutions in practical applications often requires substantial custom post-processing that motivated the development of several tools for answer set visualization \cite{DBLP:conf/iclp/CliffeVBP08,DBLP:conf/inap/KloimullnerOPT11,DBLP:journals/corr/LapauwDD15,DBLP:conf/ijcai/Bourneuf18,DBLP:journals/tplp/HahnSSS24,DBLP:journals/logcom/BertagnonG24}.
In the same spirit, ASP Chef \cite{DBLP:conf/iclp/AlvianoCR23} was introduced to streamline the construction of pipelines that combine ASP-based search and optimization tasks with other tools for filtering, aggregation, and visualization \cite{DBLP:conf/kr/AlvianoR24}. 

In previous developments, ASP Chef has demonstrated its flexibility in composing pipelines involving combinatorial search and optimization by integrating external tools through \emph{mappings}, i.e., sets of procedures that convert ASP facts into representations suitable for systems implemented in other languages. 
This methodology enabled meaningful collaborations with tools such as MiniZinc \cite{DBLP:conf/cp/NethercoteSBBDT07}, for constraint modeling and solving, and Structured Declarative Language (SDL) \cite{DBLP:conf/cilc/AlvianoDV24}, for structured problem descriptions. 
These integrations \cite{DBLP:conf/lpnmr/AlvianoGRV24,DBLP:conf/lpnmr/AlvianoGRV24} relied on defining specific interpretations of ASP facts to match the syntax and semantics required by each target system.
However, as the expressiveness and complexity of the external language increased, so did the difficulty and cost of building and maintaining the corresponding mappings.
Consequently, such efforts often focused on a limited subset of the features available in the target language, hindering full exploitation of the capabilities of the integrated tool.

In this work, we introduce a different and more scalable approach to tool integration, grounded in the use of Mustache templates (a popular web templating system; \citeNP{mittapalli2021survey}). 
Rather than building logic-based rules to obtain ASP facts that are later interpreted to interact with third-party tools, users can now define output templates using the native format of the target tool, embedding variable placeholders that are filled directly using the answer sets.
This shift in paradigm allows ASP developers to write plain templates in the language of the external tool, with minimal effort and no need for custom interpreters or adapters. 
As a result, the richness of the target system no longer translates into integration complexity. 
Instead, users are empowered to access the full feature set of the external tool by simply referring to its original documentation and writing the appropriate Mustache template. 
This template-based mechanism preserves the declarative spirit of ASP while promoting openness, modularity, and extensibility in the design of logic-based pipelines.
Thanks to Mustache, we extend ASP Chef to support the embedding of interactive views using popular JavaScript libraries such as Tabulator (for dynamic tables), Chart.js (for customizable charts), and vis.js (for networks, timelines, and 3D visualizations). 
This enhancement empowers ASP developers to generate human-readable reports, dashboards, and exploratory interfaces with minimal additional effort, all within a declarative framework.
The newly integrated libraries find application in several common and new use cases of ASP, including scheduling and data analysis.
%


\section{Background}\label{sec:background}

\paragraph{Mustache Templating System.}
Mustache is a logic-less templating system designed for generating HTML, configuration files, and other structured documents. 
Its simplicity and flexibility make it a popular choice for developers who need to separate data from presentation while avoiding complex scripting within templates.
At its core, Mustache templates use \emph{placeholders} enclosed in
\begin{asp}
    {{ ... }}
\end{asp}
to insert dynamic content.
In this work, we restrict placeholders to be \emph{variables} that get assigned values during template rendering, or \emph{sections} of the form
\begin{asp}
    {{#variable}} ... {{/variable}}
\end{asp}
that enable loops and conditionals.
Data are taken from a YAML or JSON file during template rendering.

\begin{example}
Let us consider the following Mustache template:
\begin{asp}
    The OS is {{ operating system }}. The list of users is the following:
    {{#users}}
      - {{ username }} (id {{ userid }})
    {{/users}}
\end{asp}
Applying the above template to the JSON object
\begin{asp}
    { "operating system": "Linux",
      "users": [ { "userid": 1000, "username": "alice" },
                 { "userid": 1001, "username": "bob"   } ] }
\end{asp}
renders the following text (which can be part of a Markdown document):
\begin{asp}
The OS is Linux. The list of users is the following:
 - alice (id 1000)
 - bob (id 1001)
\end{asp}
\vspace{-1.8em}
\hfill$\blacksquare$
\end{example}

\paragraph{Answer Set Programming.}
A program is a set of rules defining conditions (conjunctive bodies) under which atoms must be derived (atomic heads) or guessed (choices).
Programs are associated with zero or more answer sets, i.e., interpretations satisfying all rules and a stability condition \cite{GelfondL90}.
Programs are extended with \lstinline|#show| directives of the form
\begin{asp}
    #show $p(\overline{t})$ : $\mathit{conjunctive\_query}$.
\end{asp}
where $p$ is an optional predicate, $\overline{t}$ is a possibly empty sequence of terms, and $\mathit{conjunctive\_query}$ is a conjunction.
Answer sets of the program are projected according to the \lstinline|#show| directives.
We refer the ASP-Core-2 format for details \cite{DBLP:journals/tplp/CalimeriFGIKKLM20}.

\begin{example}\label{ex:asp}
The following program solves the K-Clique problem in ASP:
\begin{asp}
    $r_1:$  edge(X,Y) :- edge(Y,X).
    $r_2:$  {in(N) : node(N)} = K :- size(K).
    $r_3:$  :- in(X), in(Y), X < Y, not edge(X,Y).
    $r_4:$  #show (Index+1, N) : in(N), size(K), Index = #count{N': in(N'), N > N'}.
\end{asp}
Rule $r_1$ defines edge as a symmetrically closed relation (as the graph is undirected).
The choice rule $r_2$ guesses $K$ nodes (a $K$-clique candidate).
The constraint $r_3$ checks that all selected nodes are linked (a valid clique).
Finally, the \lstinline|#show| directive $r_4$ projects the answer sets over the selected nodes, indexing them according to their natural ordering.
Given \lstinline|size(3)|, and the graph
\lstinline|node(a)|,
\lstinline|node(b)|,
\lstinline|node(c)|,
\lstinline|node(d)|,
\lstinline|edge(a,b)|,
\lstinline|edge(a,c)|,
\lstinline|edge(b,c)|,
\lstinline|edge(c,d)|,
the program has one (projected) answer set, namely
\lstinline|(1, a)|,
\lstinline|(2, b)|,
\lstinline|(3, c)|.
\hfill$\blacksquare$
\end{example}

\paragraph{ASP Chef.}
An \emph{operation} $O$ is a function receiving in input a sequence of interpretations and producing in output a sequence of interpretations.
Operations may produce side outputs (e.g., a graph visualization) and accept parameters to influence their behavior.
An \emph{ingredient} is an instantiation of a parameterized operation with side output.
A \emph{recipe} is a tuple of the form \linebreak
$(\mathit{encode},\mathit{Ingredients},\mathit{decode})$, where $\mathit{Ingredients}$ is a (finite) sequence $O_1\langle{P_1}\rangle,$ $\ldots,$ $O_n\langle{P_n}\rangle$ of ingredients, and $\mathit{encode}$ and $\mathit{decode}$ are Boolean values.
If $\mathit{encode}$ is true, the input of the recipe is mapped to \mbox{\textbf{[}[\lstinline|__base64__("$s$")|]\textbf{]}}, where $s = \mathit{Base64}(s_\mathit{in})$ (i.e., the Base64--encoding of the input string $s_\mathit{in}$).
After that, the ingredients are applied one after another.
Finally, if $\mathit{decode}$ is true, every occurrence of \lstinline|__base64__($\mathit{s}$)| is replaced with (the ASCII string associated with) $\mathit{Base64}^{-1}(s)$.
Among the operations supported by ASP Chef there are 
\emph{Encode}\tuple{p, s} to extend every interpretation in input with the atom \lstinline|$p$("$t$")|, where $t = \mathit{Base64}(s)$;
\emph{Search Models}\tuple{\Pi,n} to replace every interpretation $I$ in input with up to $n$ answer sets of $\Pi \cup \{p(\overline{t}). \mid p(\overline{t}) \in I\}$;
\emph{Optimize}\tuple{\Pi,n} to replace every interpretation $I$ in input with up to $n$ optimal answer sets of $\Pi \cup \{p(\overline{t}). \mid p(\overline{t}) \in I\}$.

\begin{example}[Continuing Example~\ref{ex:asp}]\label{ex:asp-chef}
The recipe
$(\mathbf{F}, [\mathit{Search\ Models}\tuple{\{r_1, \ldots, r_4\},1}], \mathbf{F})$
addresses the K-Clique problem in a single step.
With
$(\mathbf{F}, [\mathit{Search\ Models}\tuple{\{r_1\},1}, \mathit{Search\ Models}\tuple{\{r_2, r_3\},1}, \mathit{Search\ Models}\tuple{\{r_4\},1}], \mathbf{F})$,
instead, the problem is addressed in three steps:
(i) symmetric closure of \lstinline|edge/2|;
(ii) clique search;
(iii) solution projection.
\hfill$\blacksquare$
\end{example}

\section{ASP Chef Mustache}\label{sec:Mustache}

We introduce the main linguistic constructs of our Mustache template system.
While the original system mainly deals with variables and sections to enable loops and conditionals, our system relies on ASP terms and queries.
At the core of our system there is the expansion of Mustache queries, whose results can be sorted, formatted and projected to handle duplicates (Section~\ref{sec:Mustache:core}).
More advanced constructs enable string manipulation and interpolation, the sharing of common elements among several Mustache queries, and nesting of loops and conditionals (Section~\ref{sec:Mustache:adv}).

\subsection{Core Functionalities}\label{sec:Mustache:core}

\paragraph{Mustache Queries and Expansion.}
A \emph{Mustache query} has the form \lstinline|{{ $\Pi$ }}|, where $\Pi$ is an ASP program with \lstinline|#show| directives.
Applying a Mustache query to an interpretation $I$ renders one projected answer set of $\Pi \cup \{p(\overline{t}). \mid p(\overline{t}) \in I\}$ if any, and otherwise raises an error.
Specifically, tuples of terms in the projected answer set are rendered, one per line and separating terms with ``\lstinline|, |'' (comma followed by a blank space).
A \emph{template} is a text with Mustache queries.

\begin{example}\label{ex:chef-Mustache:1}
Let $\Pi$ be the program and $I$ be the facts from Example~\ref{ex:asp}.
Applying the template
\begin{asp}
    Here is a {{ #show (K,) : size(K). }}-clique of the given graph:
    {{ $\Pi$ }}
\end{asp}
to $I$ renders
\begin{asp}
    Here is a 3-clique of the given graph:
    1, a
    2, b
    3, c
\end{asp}
Note that the rendered text depends on the order in which tuples of terms are processed.
\hfill$\blacksquare$
\end{example}

\paragraph{Shortcut Form, Sorting and Basic Formatting.}
Mustache queries are often one-liner, as \begin{asp}
    {{ #show (K,) : size(K). }}
\end{asp}
in Example~\ref{ex:chef-Mustache:1}.
Therefore, we introduce the shortcut form \begin{asp}
    {{= $p(\overline{t})$ : $\mathit{conjunctive\_query}$ }}
\end{asp}
equivalent to
\begin{asp}
    {{ #show $p(\overline{t})$ : $\mathit{conjunctive\_query}$. }}.
\end{asp}
Another shortcut is given for tuples of the form \lstinline|($t$,)| with $t$ being a number or double-quoted string, which can be equivalently written as $t$.
The processing order of tuples of terms can be specified by creating an instance of \lstinline|sort/1|, where the argument represents the index of the term used for sorting. 
A positive index indicates ascending order, while a negative index specifies descending order, using the absolute value of the index.
Ties are broken by subsequent instances of \lstinline|sort/1| (if available).
The rendering of tuples of terms can be controlled with instances of 
\lstinline|separator/1| to specify a different separator for tuples;
\lstinline|term_separator/1| to specify a different separator for terms;
\lstinline|prefix/1| and \lstinline|suffix/1| to wrap output with specified text.
As a side note, we discourage the use of guessing components (as choice rules) in Mustache queries, as they are better handled by other ASP Chef operations (e.g., \emph{Search Models} and \emph{Optimize}).

\begin{example}[Revising Example~\ref{ex:chef-Mustache:1}]\label{ex:chef-Mustache:2}
Let $I$ be the facts from Example~\ref{ex:asp} extended with \lstinline|in(a)|, \lstinline|in(b)|, \lstinline|in(c)| (i.e., the answer set of the program from Example~\ref{ex:asp} before projection).
Applying the template
\begin{asp}
    Here is a {{= K : size(K) }}-clique of the given graph: {{ 
      #show (Index + 1, N) : in(N), Index = #count{N' : in(N'), N > N'}.
      #show sort(1).
      #show separator("; ").
      #show term_separator(") ").
      #show prefix("(").
    }}.
\end{asp}
to $I$ renders
\begin{asp}
    Here is a 3-clique of the given graph: (1) a; (2) b; (3) c.
\end{asp}
Note that the order of tuples is specified in the Mustache query, and therefore the rendered text does not depend on the order in which tuples are produced by the underlying ASP solver.
\hfill$\blacksquare$
\end{example}

\paragraph{Handling Duplicates.}
ASP follows set-semantics, meaning duplicate results do not appear naturally.
On the other hand, ASP can deal with duplicate addends in sums and weak constraints thanks to an extended syntax including \emph{distinguishing terms} (i.e., terms that are used to differentiate between equal addends).
The same technique can be adopted for tuples of terms in Mustache queries:
the varadic predicate \lstinline|show/*| renders the first of its arguments (a tuple of terms), discarding all other arguments.
All of its arguments are subject to the sorting criteria specified in the Mustache query (if any), hence enabling the possibility to sort elements based on properties that are not rendered.

\begin{example}\label{ex:chef-Mustache:duplicates}
Suppose a cost is associated with each node in Example~\ref{ex:asp}:
\lstinline|cost(a,1)|,
\lstinline|cost(b,2)|,
\lstinline|cost(c,1)|,
\lstinline|cost(d,1)|.
The following weak constraint minimizes the sum of costs in the computed clique:
\begin{asp}
    :~ cost(Node,Cost), in(Node). [Cost@1, Node]
\end{asp}
Note that variable \lstinline|Node| is a distinguishing term, thanks to which the cost 1 associated with nodes \lstinline|a| and \lstinline|c| can be correctly counted twice in the computed solution.
Applying the template
\begin{asp}
    The cost is {{= S : S = #sum{Cost, Node : in(Node), cost(Node,Cost)} }} = {{
      #show show(Cost, Node) : in(Node), cost(Node,Cost).
      #show sort(2).
      #show separator(" + "). 
    }}.
\end{asp}
to the clique comprising nodes \lstinline|a|, \lstinline|b| and \lstinline|c| renders
\begin{asp}
    The cost is 4 = 1 + 2 + 1.
\end{asp}
Note that variable \lstinline|Node| is a distinguish term also in the \lstinline|#sum| aggregate and in the second Mustache query.
Also note that costs are sorted by node.
\hfill$\blacksquare$
\end{example}

\subsection{Advanced Functionalities}\label{sec:Mustache:adv}

\paragraph{Lua String @-terms.}
Mustache queries have access to some interpreted functions that ease string manipulation.
Among them, \lstinline|@string_join(sep, ...)| to concatenate two or more strings using the provided separator,
\lstinline|@string_concat(...)| as a shortcut for \lstinline|@string_join("", ...)|, and
\lstinline|@string_format(format, ...)| to format a string using the given format string and arguments.
Floating-point numbers are represented in the format \lstinline|real("NUMBER")|.

\begin{example}[Revising Example~\ref{ex:chef-Mustache:2}]\label{ex:chef-Mustache:lua-string}
The output shown in Example~\ref{ex:chef-Mustache:2} can also be obtained by the following template:
\begin{asp}
    Here is a {{= K : size(K) }}-clique of the given graph: {{ 
      #show show($\mathit{shown\_term}$, Index) :  in(N), Index = #count{N' : in(N'), N > N'}.
      #show sort(2).
      #show separator("; ").
    }}.
\end{asp}
where $\mathit{shown\_term}$ is either 
\lstinline|@string_concat("(", Index + 1, ") ", N)| or \linebreak
\lstinline|@string_format("(%d)|~\lstinline| %s", Index + 1, N)|.
\hfill$\blacksquare$
\end{example}

\paragraph{Multiline Strings and F-Strings.}
Mustache queries enrich the syntax of ASP with \emph{multiline strings} of the form \lstinline|{{"..."}}| and \emph{f-strings} of the form \lstinline|{{f"..."}}|.
A multiline string is mapped to a double-quoted string, representing new lines and double-quotes with the escape sequences \lstinline|\n| and \lstinline|\"|, respectively.
F-strings additionally introduce interpolation:
data are interpolated in f-strings using the syntax 
\lstinline|$\$${$\mathit{expression}$:$\mathit{format}$}|, where \lstinline|:$\mathit{format}$| is optional (default \lstinline|:%s| for string).
A f-string \lstinline|{{f"$\mathit{str}$"}}| is mapped to the term \lstinline|@string_format($\mathit{fmt}$, $e_1, \ldots, e_n$)|, where $\mathit{fmt}$ is the double-quoted string obtained by escaping $\mathit{str}$ and replacing interpolations with the associated formats, and $e_1, \ldots, e_n$ are the $n \geq 0$ expressions interpolated in $\mathit{str}$.

\begin{example}\label{ex:chef-Mustache:f-string}
Suppose we would like to render
\begin{asp}
    nodes: [ { id: "a", label: "a (1)", group: "in" },
             { id: "b", label: "b (2)", group: "in" },
             { id: "c", label: "c (1)", group: "in" },
             { id: "d", label: "d (1)", group: "out" } ],
\end{asp}
as part of a JSON object representing the graph and the computed clique.
We could rely on
\begin{asp}
  nodes: [{{ 
      #show show(@string_format("{id:\"%s\", label:\"%s (%d)\", group:\"out\"}", 
            X, X, C), X) : cost(X,C), not in(X).
      #show show(@string_format("{id:\"%s\", label:\"%s (%d)\", group:\"in\"}", 
            X, X, C), X) : cost(X,C), in(X).
      #show sort(2).
      #show separator(",\n").
  }}],
\end{asp}
We observe that double-quoted strings of ASP require escaping of frequent characters in JSON (the double quote char in particular).
Even worse, handling a large JSON object formatted as a single line is highly inconvenient.
Below are more convenient representations:
\begin{asp}
  nodes: [{{  #show show(@string_format({{"{ 
                id: "%s", label: "%s (%d)", group: "out"
              }"}}, X, X, C), X) : cost(X,C), not in(X).
              #show show({{f"{
                id: "$\$${X}", label: "$\$${X} ($\$${C:%d})", group: "in"
              }"}}, X) : cost(X,C), in(X).
              #show sort(2).  
              #show separator(",\n"). 
          }}],
\end{asp}
In the template above, the f-string in the second \lstinline|#show| directive essentially maps to the \linebreak \lstinline|@string_format| term in the first \lstinline|#show| directive (modulo the value of \lstinline|group|).
\hfill$\blacksquare$
\end{example}

\paragraph{Persistent Queries.}
Common elements of different Mustache queries in a template can be stored in a \emph{persistent array}, initially empty, by using the expression 
\begin{asp}
    {{* $\Pi$ }}
\end{asp}
where $\Pi$ is an ASP program with \lstinline|#show| directives;
or using the shortcut form 
\begin{asp}
    {{+ $p(\overline{t})$ : $\mathit{conjunctive\_query}$ }}
\end{asp}
equivalent to
\begin{asp}
    {{* #show $p(\overline{t})$ : $\mathit{conjunctive\_query}$. }}.
\end{asp}
The content of the persistent array is prepended to the tuples of terms and atoms obtained by evaluating subsequent Mustache queries in the template.
The persistent array can be reset using the Mustache expression \lstinline|{{-}}|.

\begin{example}[Continuing Example~\ref{ex:chef-Mustache:f-string}]\label{ex:chef-Mustache:persist}
Targeting the rendering of JSON objects, the separator is likely always \lstinline|",\n"|.
Instead of including the associated \lstinline|#show| directive in all Mustache queries, a template could start with the Mustache expression \lstinline|{{+ separator(",\n")|~\lstinline| }}|.
\hfill$\blacksquare$
\end{example}

\paragraph{Multi-Stage Expansion.}
Mustache queries enable loops and conditionals.
In some cases, nesting of Mustache queries is convenient for evaluating conditionals within loop elements or for executing inner loops.
A multi-stage template expansion repeatedly processes Mustache queries until no further expansions remain.

\begin{example}[Continuing Examples~\ref{ex:chef-Mustache:f-string}--\ref{ex:chef-Mustache:persist}]\label{ex:chef-Mustache:multi-stage}
As already observed, the multiline string and the f-string in Example~\ref{ex:chef-Mustache:f-string} only differ in the value of \lstinline|group|.
Nesting a conditional within a single loop over nodes seems natural in this case.
Assuming \lstinline|sort(2)| and \lstinline|separator(",\n")| are in the persistent array, we can use
\begin{asp}
    nodes: [ {{= show({{f"
        { 
          id: "$\$${X}", 
          label: "$\$${X} ($\$${C:%d})", 
          group: {{= "out" : not in($\$${X}) }}{{= "in" : in($\$${X}) }}
        }
      "}}, X) : cost(X,C) 
    }} ],
\end{asp}
Note that the first expansion renders (four inner objects like) the following:
\begin{asp}
    nodes: [ { id: "a", label: "a (1)", 
               group: {{= "out" : not in(a) }}{{= "in" : in(a) }} }, $\cdots$ ],
\end{asp}
Conditionals within each object are evaluated in the second stage to obtain the \lstinline|group| values.
\hfill$\blacksquare$
\end{example}

\section{JSON-Based Frameworks Integration}\label{sec:frameworks}

Several frameworks can be configured using JSON objects.
To further ease their integration in ASP Chef, we rely on \emph{Relaxed JSON} (\url{https://www.relaxedjson.org/}), hence accepting a more permissive syntax.
The new operations presented in this section have parameters \tuple{p,m}, where $p$ is a predicate and $m$ is a Boolean.
Each atom $p(s)$ in each input interpretation $I$ produces a side output according to the JSON object rendered by applying the template $\mathit{Base64}^{-1}(s)$ to $I$, using multi-stage expansion if $m$ is $\mathbf{T}$.

\begin{figure}
    \centering
    \includegraphics[width=0.2\textwidth]{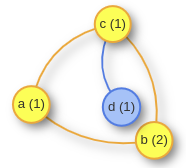}
    \hspace{1.5em}
    \includegraphics[width=0.25\textwidth]{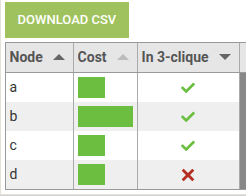}
    \hspace{1.5em}
    \includegraphics[width=0.45\textwidth]{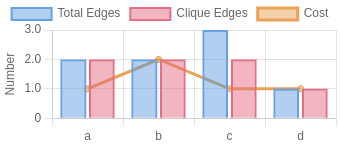}
    \caption{
        Side outputs associated with data from Example~\ref{ex:chef-Mustache:duplicates}:
        graph with highlighted clique obtained with the \emph{@vis.js/Network} operation;
        table showing node costs and computed clique obtained with the \emph{Tabulator} operation;
        chart reporting statistics obtained with the \emph{Chart.js} operation.
        A recipe showcasing these examples is available at \protect\url{https://asp-chef.alviano.net/s/ICLP2025/running-example}.
    }
    \label{fig:tabulator}
\end{figure}

\paragraph{@vis.js/Network.}
The Network module in vis.js (\url{https://visjs.org/}) is a powerful JavaScript library for visualizing dynamic and interactive networks (graphs). 
Users can customize colors, shapes, labels, and border styles of nodes and edges.
The module includes a physics-based layout engine, and supports real-time interactivity.
Additionally, grouping and clustering allow for efficient visualization of large datasets by aggregating related nodes. 
The library also supports hierarchical layouts and directional edges with arrows.
We extended ASP Chef with the \emph{@vis.js/Network} operation.

\begin{example}[Display data from Example~\ref{ex:chef-Mustache:duplicates} as a graph]
The following template renders the graph shown in Figure~\ref{fig:tabulator}:
\begin{asp}
{ data: {
    nodes: [ {{= show({{f"
        { 
          id: "$\$${X}", 
          label: "$\$${X} ($\$${C:%d})",
          group: {{= "out" : not in($\$${X}) }}{{= "in" : in($\$${X}) }}
        }
      "}}, X) : cost(X,C)
    }} ],
    edges: [
      {{= {{f"{ from: "$\$${X}", to: "$\$${Y}" }"}} : edge(X,Y), X<Y, in(Y) }}
      {{= {{f"{ from: "$\$${Y}", to: "$\$${X}" }"}} : edge(X,Y), X<Y, not in(Y) }} 
    ] 
  },
  options: {
    nodes: { shape: "circle", borderWidth: 2, shadow: true },
    edges: { width: 2, shadow: true },
    groups: { 
      "in": { color: { background: "yellow", border: "orange"} } }
    }
}
\end{asp}
Note that nodes in the \lstinline|"in"| group (and edges within the clique) are highlighted in yellow.
\hfill$\blacksquare$
\end{example}

\paragraph{Tabulator.}
Tabulator (\url{https://tabulator.info/}) is a powerful JavaScript library that provides a wide range of customization options to create interactive tables from various sources, including JSON. 
In the context of ASP Chef, Tabulator can be used to render tables from input models, taking advantage of Mustache queries to configure the table and bind data from ASP facts as needed.
Specifically, we extended ASP Chef with the \emph{Tabulator} operation, and enriched the JSON configuration to accept the \lstinline|download| property, allowing users to add buttons for exporting the data in various formats such as CSV, JSON, or Excel.

\begin{example}[Display data from Example~\ref{ex:chef-Mustache:duplicates} as a table]
The following template renders the table shown in Figure~\ref{fig:tabulator}:
\begin{asp}
{ data: [{{= {{f"
      { 
        node: "$\$${Node}", 
        cost: $\$${Cost * 100 / Max},
        in: {{= "true": in($\$${Node}) }}{{= "false": not in($\$${Node}) }},
      },
    "}} : cost(Node, Cost), Max = #max{Cost' : cost(_, Cost')}
  }}],
  columns: [ { title: "Node", field: "node" },
             { title: "Cost", field: "cost", formatter: "progress" },
             { title: "In {{= K : size(K) }}-clique", field: "in", 
               hozAlign: "center", formatter: "tickCross" } ],
  initialSort: [ { column: "node", dir: "asc" }, { column: "in", dir: "desc" } ],
  download: [ { color: "success", format: "csv", options: { delimiter: "\t" } } ]
}
\end{asp}
Note that \lstinline|data| is a list of objects whose properties are obtained interpolating ASP terms.
Costs are shown as a (progress) bar, and membership in the computed clique using ticks.
Rows are sorted by membership in the clique (descending), and node name (ascending).
Finally, the user can export data as comma-separated values (CSV) by clicking a button.
\hfill$\blacksquare$
\end{example}

\paragraph{Chart.js.}
Chart.js (\url{https://www.chartjs.org/}) is a lightweight yet powerful JavaScript library for creating interactive and customizable charts. 
It supports multiple chart types, including line, bar, pie, radar, and scatter plots, making it versatile for various data visualization needs. 
Its JSON configuration can easily customize styles, adjusting colors, fonts, and tooltips.
We extended ASP Chef with the \emph{Chart.js} operation.

\begin{example}[Display statistics about Example~\ref{ex:chef-Mustache:duplicates}]
The following template renders the mixed chart (bar plots and line) shown in Figure~\ref{fig:tabulator}:
\begin{asp}
{{+ sort(2) }}
{ type: "bar",
  data: {
    labels: [{{= {{f"$\$${Node}"}} : node(Node) }}],
    datasets: [{
      label: "Total Edges", borderWidth: 2,
      data: [{{= show(V,N) : node(N), V = #count{N' : edge(N,N')} }}],
    }, {
      label: "Clique Edges", borderWidth: 2,
      data: [{{= show(V,N) : node(N), V = #count{N' : edge(N,N'), in(N')} }}],
    }, {
      type: "line", label: "Cost",
      data: [{{= show(Value, Node) : cost(Node, Value) }}],
    }] },
  options: { 
    scales: {
      y: { beginAtZero: true,  title: { display: true, text: "Number" } } }
    }
}
\end{asp}
The template defines three datasets, two bars and one line, using the \lstinline|show/*| predicate to handle duplicate values.
Also note that values are sorted by node (the second argument in \lstinline|show/*|).
\hfill$\blacksquare$
\end{example}

\paragraph{Other Frameworks}
We extended ASP Chef with other frameworks, providing alternatives to build charts and images.
The \emph{@vis.js/Timeline} operation is specialized for temporal data.
The library supports custom styling and grouping of events, making it ideal for project management, historical data visualization, and scheduling applications. 
The \emph{@vis.js/Graph3D} operation can create interactive 3D visualizations of data, making it ideal for representing mathematical functions, scientific data, and geographical information.
Data points are rendered on a 3D plane, either as surface plots or scatter plots.
The module provides intuitive controls for zooming, rotating, and panning, allowing users to explore complex datasets from different angles.
Customization options include color gradients, axis scaling, and grid styling, enabling precise data representation. 
The \emph{ApexCharts} (\url{https://apexcharts.com/}) operation integrates a modern, highly customizable JavaScript charting library designed for creating interactive, responsive, and performant visualizations.
It supports a wide range of chart types, including line, bar, area, pie, radar, heatmaps, and mixed charts, making it suitable for business intelligence dashboards, financial data analysis, and real-time monitoring. 
Interactivity is a key strength, with built-in support for tooltips, zooming and panning.
The \emph{Fabric.js} (\url{https://fabricjs.com/}) operation integrates a powerful and flexible JavaScript library for working with HTML5 canvas, enabling rich interactive graphics, image manipulation, and object-based drawing.
It simplifies complex vector graphics operations by providing an intuitive API for creating and managing shapes, images, text, and paths. 
One of its standout features is object-based manipulation, allowing users to move, scale, rotate, and group elements directly on the canvas.

\section{Use Cases}\label{sec:usecases}

\begin{figure}
    \centering
    \includegraphics[width=0.65\textwidth]{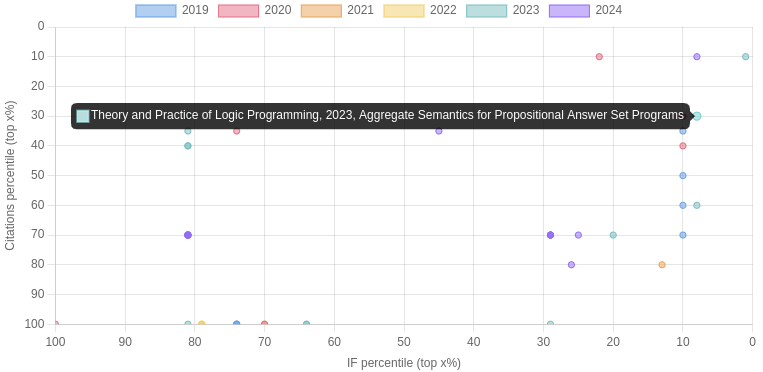}
    \hspace{1em}
    \includegraphics[width=0.3\textwidth]{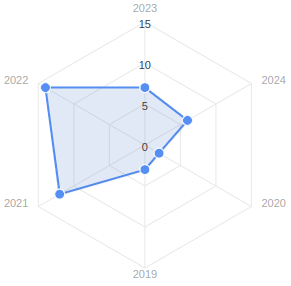}
    \caption{
        VQR data analysis with \emph{Chart.js} and \emph{ApexChart}
    }
    \label{fig:vqr}
\end{figure}

\paragraph{Data Analysis.}
Italian VQR (Valutazione della Qualità della Ricerca) is the national framework used to assess the quality and impact of research outputs through bibliometric indicators such as citation percentiles and journal impact factors.
ASP finds a natural application in this context to optimize the selection of articles avoiding conflicts (see \url{https://asp-chef.alviano.net/s/VQR2024/assegna-paper}).
The newly integrated libraries add several useful capabilities to analyze input data and results.
Figure~\ref{fig:vqr} shows a scatter plot obtained with \emph{Chart.js}, where each point combines the percentiles on citations and impact factor to provide an insightful overview of research performance, and a radar chart from \emph{ApexChart} to overview the number of articles per year included in the analysis.
An interactive recipe starting from CSV and including a \emph{Tabulator} representation is available at \url{https://asp-chef.alviano.net/s/ICLP2025/vqr}.

\begin{figure}
    \centering
    \includegraphics[width=0.13\textwidth]{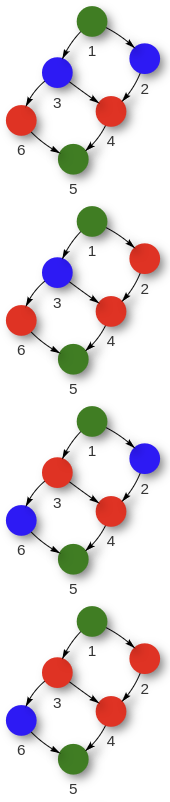}
    \includegraphics[width=0.13\textwidth]{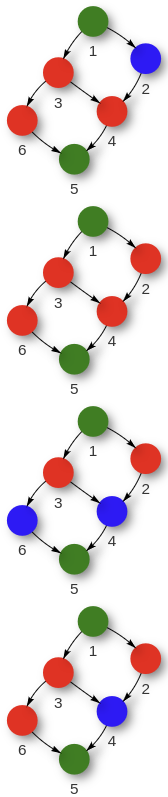}
    \hspace{2em}
    \includegraphics[width=0.65\textwidth]{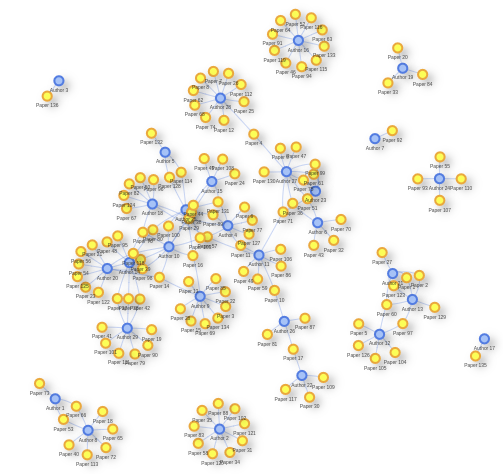}
    \caption{
        Blocking sets and authorship graphs obtained with \emph{@vis.js/Network}
    }
    \label{fig:graphs}
\end{figure}

\paragraph{Graph Analysis.}
The concept of \emph{blocking sets} is also known as vertex cuts or vertex separators in graph theory: given a start and an end vertex, a blocking set is a set of vertices (without the start and end vertices) such that the end vertex is not reachable from the start vertex without passing a member of the blocking set. The identification of blocking sets has numerous applications, for example in VLSI design and cybersecurity.
Using \emph{@vis.js/Network}, ASP Chef can display blocking sets of a given directed graph, using the same layout for all computed solution to ease their understanding.
A recipe is available at
\url{https://asp-chef.alviano.net/s/ICLP2025/blocking-sets}.
For a larger example, 
\url{https://asp-chef.alviano.net/s/ICLP2025/vqr2}
shows an authorship graph from which different communities of authors are easily spotted and independently optimized to select articles for the Italian VQR.
Figure~\ref{fig:graphs} reports networks obtained with \emph{@vis.js/Network} in these contextes. For blocking sets, the start and end vertices are green, vertices in the blocking sets are red.

\begin{figure}
    \centering
    \includegraphics[width=0.8\textwidth]{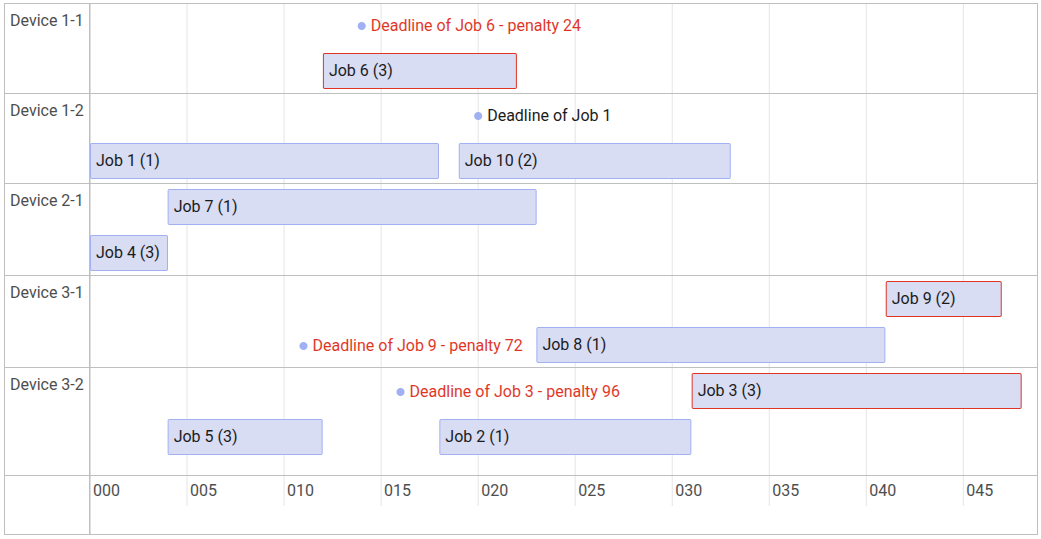}
    \hspace{.1em}
    \includegraphics[width=0.15\textwidth]{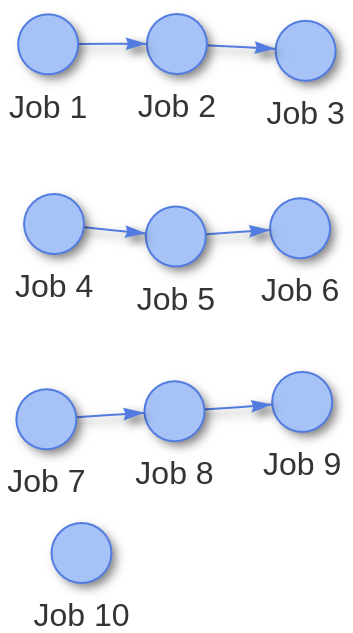}
    \caption{
        Computed scheduling and job dependencies shown with \emph{@vis.js/Timeline} and \emph{@vis.js/Network} 
    }
    \label{fig:scheduling}
\end{figure}

\paragraph{Planning and Scheduling.}
Incremental Scheduling is a complex problem that involves allocating jobs to specific devices with replicas, while considering deadlines, dependencies, and importance levels.
The goal is to minimize the penalty for missing deadlines and the overall completion time.
To visualize this intricate process, we utilize \emph{@vis.js/Timeline} to create a dynamic and interactive schedule. 
The timeline ingredient showcases the computed schedule, highlighting in red the penalties incurred for missing deadlines. 
This visualization provides a clear and concise overview of the scheduling process, allowing users to easily identify areas of improvement and optimize their scheduling strategy.
Additionally, a \emph{@vis.js/Network} visualization is used to illustrate the dependencies between jobs, providing a comprehensive understanding of the scheduling problem.
Figure~\ref{fig:scheduling} reports the two visualizations obtained with the recipe at \url{https://asp-chef.alviano.net/s/ICLP2025/incremental-scheduling}.

\begin{figure}
    \centering
    \includegraphics[width=0.25\textwidth]{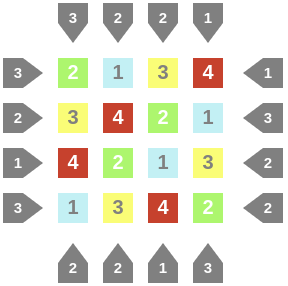}
    \hspace{1em}
    \includegraphics[width=0.35\textwidth]{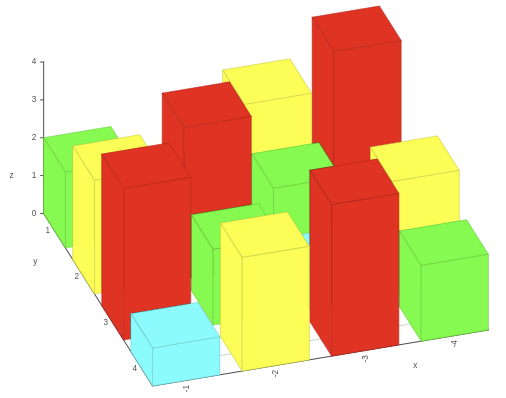}
    \hspace{1em}
    \includegraphics[width=0.3\textwidth]{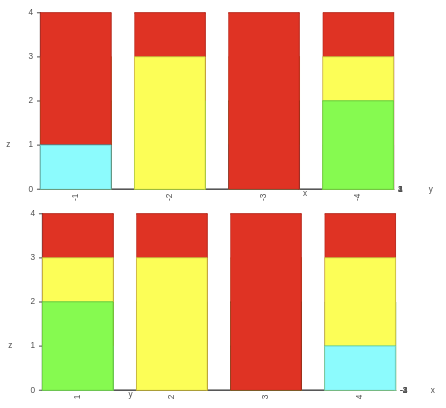}
    \caption{
        Skyscrapers solution shown using \emph{Fabric.js} and \emph{@vis.js/Graph3d} (from different angles)
    }
    \label{fig:skyscraper}
\end{figure}

\paragraph{2D and 3D Solution Visualization.}
In the Skyscrapers puzzle the goal is to determine the height of skyscrapers in a grid, given clues on the number of visible skyscrapers in some directions (\url{https://www.puzzle-skyscrapers.com/}).
A recipe addressing the puzzle is available at \linebreak \url{https://asp-chef.alviano.net/s/ICLP2025/skyscrapers}, and
produces the visualizations shown in Figure~\ref{fig:skyscraper}.
The 3D visualization is particularly interesting here, as the user can use the mouse to control the camera angle to obtain an immersive experience.

\section{Extending ASP Chef with New Operations}\label{sec:impl-details}

A key strength of ASP Chef lies in its deliberately simple and modular architecture, which enables users to extend the system by adding new operations with minimal effort.
Rather than modifying core infrastructure or registering components manually, operations are implemented as self-contained (and self-registering) \lstinline|.svelte| files that become instantly usable within the system.
This design encourages experimentation, facilitates integration of third-party libraries, and lowers the barrier for community contributions.
To implement a new Mustache-based visualization operation (such as a charting or table component) developers typically create two files within \lstinline|src/lib/operations|:
(i)
a file named \lstinline|<Operation>.svelte| that defines UI controls for parameters of the operation (e.g., \lstinline|predicate| and \lstinline|multi-stage|), and
(ii)
a companion file named \lstinline|+<Operation>.svelte| that imports and renders the third-party library, feeding it the result of the Mustache template as configuration.

For example, for the \emph{ApexCharts} operation, the main \lstinline|ApexCharts.svelte| file contains
\begin{asp}
<Operation {id} {operation} {options}>
  <Input type="text" bind:value={options.predicate} on:input={edit} />
  <Button outline="{!options.multistage}" on:click={() => { 
      options.multistage = !options.multistage; edit(); }}>Multi-Stage</Button>
  {#each models.flatMap(model => 
              model.filter(atom => atom.predicate === options.predicate)) as cfg}
    <ApexCharts part={model} cfg_atom={cfg} multistage={options.multistage} />
  {/each}
</Operation>
\end{asp}
This snippet instantiates the operation with UI controls to select the ASP predicate that encodes the chart configuration and to toggle options such as multi-stage.
The \lstinline|#each| loop adds instances of \lstinline|+ApexCharts.svelte| for each instance of the selected predicate.
%
%
%
The companion file \lstinline|+ApexCharts.svelte| integrates the ApexCharts library and renders the chart:
\begin{asp}
<script>
    import ApexCharts from 'apexcharts'; import {Utils} from "$\$$lib/utils";
    import {Base64} from "js-base64";    import {onMount} from "svelte";

    export let part, cfg_atom, multistage;
    let chart;
    onMount(async () => {
        const content = Base64.decode(cfg_atom.terms[0].string);
        const expanded_content = await Utils.expand_mustache_queries(
                                                part, content, multistage);
        const configuration = Utils.parse_relaxed_json(expanded_content);
        await (new ApexCharts(chart, configuration)).render();
    });
</script>
<div class="chart" bind:this={chart}></div>
\end{asp}
This fragment decodes the Mustache template in the configuration atom, expands it using the current interpretation (\lstinline|part|) as context, parses it as JSON, and invokes the rendering engine.

Despite their functionality, these files are concise:
(i) \lstinline|ApexCharts.svelte| is 73 lines;
(ii) \lstinline|+ApexCharts.svelte| is 44 lines.
Similar patterns are followed for other operations:
\emph{Chart.js} (73 + 47 lines) replicates the same logic with a different charting library;
\emph{Tabulator} (74 + 70 lines) extends slightly to add features like export buttons.

\section{Related Work}\label{sec:rw}

Several tools and frameworks supporting the development of ASP have been introduced in the literature, among them 
Integrated Development Environments (IDEs) such as \textsc{ASPIDE}~\cite{DBLP:conf/lpnmr/FebbraroRR11}, \textsc{SeaLion}~\cite{DBLP:journals/tplp/BusoniuOPST13}, and \textsc{LoIDE}~\cite{DBLP:journals/ki/CalimeriGPRR18}.
Unlike these tools, ASP Chef is designed as a platform for experimenting with ASP solutions and their integration with external tools.
ASP Chef is powered by \textsc{clingo-wasm} and enables the development of shareable examples and interactive demos, a feature especially suited to scientific communication, as shown in \cite{DBLP:conf/cilc/Costantini024}.
Several research efforts have explored the modular use of ASP through microservices or pipeline-based compositions~\cite{DBLP:journals/aicom/CalimeriI06,DBLP:conf/cilc/CostantiniG18,DBLP:journals/tplp/CabalarFL20,DBLP:journals/ijimai/CostantiniGL21,DBLP:journals/algorithms/CabalarFSW23}. 
These works resonate with the guiding principles of ASP Chef, which promotes the decomposition of ASP workflows into manageable and reusable components.
Prior versions of ASP Chef supported integration with external systems, such as MiniZinc and Structured Declarative Language (SDL), by exchanging ASP facts.
However, this fact-based approach often incurred non-trivial implementation costs to interpret ASP data in the target language, especially when the target system had a rich and complex syntax.
To address this limitation, the present work introduces a more direct and extensible mechanism for integration, based on Mustache templates. 
Instead of interpreting facts, the user defines templates that interpolate ASP data into the desired target format or language, facilitating smoother and more expressive interactions with external libraries and systems. 
This enables seamless integration of advanced visualization and reporting tools, without needing dedicated wrappers or converters.

A first embryonic version of templated queries was introduced in our earlier work \emph{ASP Chef Chats with Large Language Models} \cite{ijcai-llms}, where a restricted form of Mustache query was proposed to synthesize structured Markdown snippets from ASP results and feed them into LLM prompts.
In that version, the output was a projection of matching atoms, formatted via special atoms like \lstinline|ol/1|, \lstinline|ul/1|, \lstinline|th| and \lstinline|tr|. The goal was to enable Markdown generation suitable for LLM consumption, with Base64-encoded answers retrieved from predicates such as \lstinline|__base64__/1|.
In contrast, the current work generalizes this idea into a full-featured templating system, decoupled from LLM usage and capable of targeting arbitrary external libraries.
By adopting standard Mustache templates and extending them with ASP-aware data injection, we shift from producing static Markdown to dynamically generating JSON configurations, HTML fragments, and other consumable formats suitable for integration with a wide range of front-end and visualization frameworks. 

Visual representation of ASP results has been studied through tools such as \textsc{ASPViz}~\cite{DBLP:conf/iclp/CliffeVBP08}, \textsc{IDPD3}~\cite{DBLP:journals/corr/LapauwDD15}, and \textsc{Kara}~\cite{DBLP:conf/inap/KloimullnerOPT11}, all of which rely on ASP facts to describe the visual layout of answer sets. 
In the same spirit, more recent approaches like \textsc{clingraph}~\cite{DBLP:conf/lpnmr/HahnSSS22} and \textsc{ASPECT}~\cite{DBLP:conf/cilc/BertagnonGZ23} offer exportable, high-quality visualizations suited for publication in \LaTeX.
While these systems provide expressive rendering capabilities, they generally depend on external installations and do not focus on browser-native interaction. 
Moreover, their development and maintenance must face the complexity of the underlying rendering engines.
The visualization operations introduced in this work, based on Tabulator, Chart.js, and vis.js, complement these efforts by enabling fully browser-based, interactive visualizations such as scatter plots, timelines, and network diagrams. 
Building on our experience with the previously developed \emph{Graph} operation~\cite{DBLP:conf/kr/AlvianoR24}, we found that using Mustache templates to implement the \emph{@vis.js/Network} and \emph{Fabric.js} operations not only significantly reduced development time but also led to more powerful and expressive functionalities.

A particularly relevant application of ASP-based visualization in complex domains is presented by Gebser et al.~\citeyear{DBLP:journals/tplp/GebserOOS18}, who propose an ASP-based framework for experimenting with robotic intra-logistics domains. 
Their work highlights the value of integrating high-level reasoning with sophisticated visualization and execution environments, and shows how ASP can serve as a flexible backbone for modeling, simulation, and visualization in complex logistical settings. 
ASP Chef shares this ambition but places stronger emphasis on interactive, browser-native visualizations that support rapid prototyping and deployment via web technologies.
Finally, we mention \textsc{clinguin}~\cite{DBLP:journals/corr/abs-2502-09222} for developing graphical user interfaces directly in ASP.
Future applications of Mustache in ASP Chef may explore convergence points between these systems, particularly in the direction of form-based input via frameworks like SurveyJS.

\section{Conclusion}\label{sec:conclusion}

We presented an extension of ASP Chef that uses Mustache templates and popular JavaScript libraries to support the creation of interactive, data-driven views from ASP results. 
By simplifying the integration of external visualization tools, our approach enables developers to build rich, user-friendly interfaces while staying within the declarative paradigm.

Beyond its functionality, a key strength of ASP Chef is its simple and modular design. 
New Mustache-based operations can be added with minimal effort by creating a \lstinline|.svelte| file that defines UI parameters and links to a third-party library using the configuration generated from the expanded template. 
These operations are immediately usable in recipes, without the need for manual registration or system modification. 
This plugin-like architecture lowers the barrier to contribution and invites experimentation. 
In future iterations, we plan to provide formal guidelines to support community-driven extensions and integrations.

Future work will explore the use of Mustache templates to integrate frameworks for user interface definition, particularly form-based interactions. 
This would allow ASP developers not only to present data, but also to collect and structure user input within a seamless and declarative pipeline.
In addition, we envision integrating Controlled Natural Language (CNL) processing capabilities to allow domain experts to write specifications in a restricted natural language that can be automatically translated into ASP code \cite{DBLP:journals/tplp/CarusoDMMR24}. 
The combination of Mustache templating, structured input collection, and CNL2ASP translation would broaden the accessibility of ASP Chef, making it easier for non-programmers to contribute to the knowledge engineering process.


\section*{Acknowledgments}
This work was supported 
by the Italian Ministry of University and Research (MUR) 
    under PRIN project PRODE ``Probabilistic declarative process mining'', CUP H53D23003420006,
    under PNRR project FAIR ``Future AI Research'', CUP H23C22000860006,
    under PNRR project Tech4You ``Technologies for climate change adaptation and quality of life improvement'', CUP H23C22000370006, and~    under PNRR project SERICS ``SEcurity and RIghts in the CyberSpace'', CUP H73C22000880001;
by the Italian Ministry of Health (MSAL)
    under POS projects \linebreak CAL.HUB.RIA (CUP H53C22000800006) and RADIOAMICA (CUP H53C22000650006);
by the Italian Ministry of Enterprises and Made in Italy
    under project STROKE 5.0 (CUP \linebreak B29J23000430005);
    under PN RIC project ASVIN ``Assistente Virtuale Intelligente di Negozio'' (CUP B29J24000200005);
and by the LAIA lab (part of the SILA labs). 
This research was also funded in part by the Austrian Science Fund (FWF) within projects 10.55776/COE12 and 10.55776/PIN8782623.
Mario Alviano is member of Gruppo Nazionale Calcolo Scientifico-Istituto Nazionale di Alta Matematica (GNCS-INdAM).

\bibliographystyle{acmtrans}
\bibliography{bibtex}

\label{lastpage}
\end{document}